# Dilute Zn alloying in biodegradable Mg wires: microstructure, mechanical performance, and degradation behavior

Jiří Ryjáček [a], Leonard Hlodák [a], Jiří Liška [a,b], Jan Pinc [b], Tomáš Herma [c], Karel Tesař [a,*]

[a] Department of Materials, Faculty of Nuclear Sciences and Physical Engineering, Czech Technical University in Prague, Trojanova 13, Prague 120 00, Czech Republic
[b] FZU - Institute of Physics of the Czech Academy of Sciences, Na Slovance 2, Prague 182 21, Czech Republic
[c] Department of Plastic Surgery, Kralovske Vinohrady University Hospital and Third Faculty of Medicine, Charles University, Šrobárova 50, Prague 100 34, Czech Republic

ABSTRACT

Dilute Mg-Zn wires are of interest for biodegradable small-bone fixation as magnesium degradation can support bone healing, while low zinc additions may provide biological benefit without compromising biocompatibility. The influence of Zn content below the room-temperature solubility limit was assessed in Mg-Zn wires intended for resorbable implant applications. Mg-0.4Zn, Mg-0.6Zn, Mg-0.8Zn, and Mg-1.5Zn alloys were processed by single-step direct hot extrusion into thin wires and characterized via correlative microstructural analysis, tensile testing, bending experiments, and in vitro degradation. All compositions achieved a recrystallized equiaxed grain size of 5.0–5.9 μm and exhibited ultimate tensile strengths of 246–256 MPa with elongations of 23–28 %. Zn content had only a limited effect on grain size and ultimate tensile strength, whereas the 0.2 % proof stress was higher for Mg-0.4Zn and Mg-0.6Zn, which exhibited a pronounced sharp yield point. Bending of Mg-0.6Zn and Mg-1.5Zn was governed mainly by extrusion texture and preserved reversible plasticity through twinning and detwinning. Simulated body fluid caused rapid, localized degradation and loss of mechanical integrity within 7 days, whereas an application-relevant medium (DMEM + FBS) resulted in slower degradation and higher retained strength. These findings support dilute Mg-Zn wires as a material platform for future resorbable bone fixation devices.

## 1. Introduction

Resorbable magnesium alloys with small zinc additions are promising implant materials for bone-support applications, where the active roles of Mg and Zn in bone metabolic processes may provide a substantial benefit for healing [1–3], although some effects in the later healing stages are still being explored [4]. At the same time, the absence of secondary removal surgeries benefits both patients and healthcare systems. Especially dilute Mg-Zn alloys, with Zn concentrations below the room-temperature solubility limit of ~1.6 wt% [5], are of interest for implant applications, as the absence of micrometer-scale Mg-Zn-based phases reduces the likelihood of localized corrosion and, consequently, the risk of premature loss of mechanical integrity in final medical devices [6]. Furthermore, low concentrations of alloying elements may limit potential safety risks associated with such resorbable implants, including hypersensitivity [7], toxicity [8], and insufficiently explored long-term effects [9,10].

Mg-based resorbable wires, as one much-needed implant class, have been investigated in numerous studies, including works focused on direct extrusion and wire drawing [11–13], degradation behavior and the loss of mechanical integrity under immersion conditions [14,15], and the in vitro and *in vivo* performance [16–18]. In addition, recent machine-learning approaches have been applied to accelerate the development of lean and dilute Mg alloys and to link composition, processing, and properties, which may provide useful guidance for future Mg-Zn wire design [19,20]. When the final implant design is considered, Mg-based wires have been applied in various forms, including stents [21], staples [22], cerclage wires [18] and wire-tendon sutures [23].

Although such wires constitute a promising innovation in orthopedic and trauma surgery [24], the mechanical and degradation behavior of small and thin-walled Mg-based implants is still not fully understood. One open question is whether wires with low Zn alloying can maintain favorable reversible twinning as a mechanism of plastic deformation during repeated bending, as observed in pure Mg [11]. Other such questions relate to the effects of low Zn alloying on the mechanical and degradation properties of the resulting wires. The motivation behind this study is that dilute binary Mg-Zn alloys could provide a valuable platform for effective and biologically safe implants, including those intended for pediatric patients, provided that they show adequate performance. This is particularly relevant because the required mechanical parameters of implants for children are generally lower than those for adult patients, while the safety requirements may be more demanding due to the developing organism. At the same time, such alloys would still allow further microalloying with rare-earth or other elements whose effects on the developing body of children remain more debatable, especially for adult applications, where the risks associated with adverse effects of released metallic ions are generally lower [10].

* Corresponding author.
*E-mail address:* Karel.Tesar@fjfi.cvut.cz (K. Tesař).

In this study, dilute binary Mg-Zn wires with Zn concentrations of 0.4, 0.6, 0.8, and 1.5 wt% were prepared by direct hot extrusion and comparatively evaluated with respect to their microstructure, mechanical properties, corrosion behavior, and degradation-induced loss of mechanical integrity. Particular attention was paid to the effect of Zn content below the room-temperature solubility limit on the tensile properties and bending-related deformation behavior of the wires, including the possible preservation of twinning-detwinning-assisted reversible plasticity. Based on the initial screening, the most promising alloy was selected for more detailed in vitro degradation testing, including the analysis of corrosion products and residual mechanical performance after immersion in simulated physiological media. Together, these analyses were intended to assess whether dilute binary Mg-Zn wires can provide a mechanically reliable and biologically conservative material platform for future resorbable implants in small-bone fixation.

## 2. Materials and methods

### 2.1. Manufacturing

Cast ingots of the Mg-0.4Zn, Mg-0.6Zn, Mg-0.8Zn, and Mg-1.5Zn alloys were supplied by Crown Metals CZ s.r.o. All four compositions were prepared from the same magnesium base material, with only the Zn content adjusted to reach the target levels. The chemical compositions measured by the supplier using spark optical emission spectrometry are given in Table 1.

Cylindrical samples, each 7 mm in diameter and 22 mm in height, were cut from each ingot by electrical discharge machining (EDM) and subsequently machined to final dimensions of 6 mm in diameter and 20 mm in height to remove any brass-wire contamination introduced during EDM. Prior to extrusion, the billets were pickled in 7 % Nital for 2 min and rinsed in ethanol. Direct hot extrusion was performed at 300 °C with an extrusion ratio of 400:1 and a constant ram speed of 0.2 mm/s. During the process, high-temperature paste GLEIT-μ HP 505 was used as a lubricant. It should be noted that temperature control during extrusion was not precise, as the temperature could be monitored only between successive runs and not under active load. This process yields wires with a final diameter in the 290–300 μm range, due to minor die geometry changes during the extrusion process and wire thermal contraction as the wires cool down to room temperature.

### 2.2. Microstructure characterization

For electron back-scattered diffraction (EBSD) analysis, wires were embedded in Technovit 5000 (Kulzer Technik). After curing, the samples were ground and polished using standard metallographic procedures. The final polishing step was performed using a colloidal silica suspension OP-S (Struers) with subsequent etching for 1 s in a solution with the following composition (vol. %): 15 % $CH_3COOH$, 5 % $HNO_3$, 20 % $H_2O$ and 60 % ethanol.

EBSD analysis was conducted on a JSM-IT500HR (JEOL) scanning electron microscope (SEM) using a Velocity EBSD camera (EDAX) at an acceleration voltage of 30 kV and a step size of 200–300 nm. Diffraction patterns were processed by spherical indexing and analyzed using the TSL OIM Analysis 9 software. For a detailed discussion on the effect of the spherical indexing options used and the step size used, see Section 2 in the Supporting Information. EBSD was used to determine pole figures (PF), inverse pole figure (IPF) maps, grain orientation spread (GOS) maps, crystal direction (CD) maps, and kernel average misorientation (KAM) maps. From the EBSD data, the number fraction and average value of the equivalent circle diameter were determined for each wire composition. The equivalent circle diameter of each grain was calculated from its measured area $A$ as $d = 2\sqrt{A/\pi}$, i.e., the diameter of a circle of equivalent area, and is hereafter referred to as the grain size in this manuscript.

Three-dimensional visualization of the wire cross-section prior to etching, together with volumetric analysis of particles, was performed using a computed tomography Zeiss Xradia 610 Versa system (μCT) with a voxel size of 470 nm. The acquired data were processed using Dragonfly software (v. 2025.1). The dataset was segmented to distinguish the wire, particles, and background using histographic segmentation, followed by visual inspection and determination of the volume fraction of particles relative to the wire. It should be noted that partial error in the measured oxide content may arise from segmentation artefacts, such as residual noise, partial volume effects, or insufficient contrast between phases, which may lead to voxel misclassification. However, this error is expected to be minimal due to the limited number of visible artefacts in the scans. Impurities were characterized using secondary electrons (SE) in SEM using an accelerating voltage of 15 kV and 20 kV, with chemical analysis conducted via energy-dispersive spectroscopy (EDS) using Octane Elite Super (EDAX) detector. Image analysis on SEM-SE images was used to calculate the area fraction of impurities in the wire cross-section.

### 2.3. Mechanical testing

Prior to tensile testing, Mg-0.4Zn, Mg-0.6Zn, Mg-0.8Zn, and Mg-1.5Zn wire segments were pickled in a 7 wt% nital solution for approximately 30 s, removing at least 10 μm from the surface to eliminate surface defects introduced during processing, following the approach described in [18]. The wire ends were sandwiched between pairs of mounting tabs and bonded with cyanoacrylate adhesive. The mounting tabs were fabricated from polylactic acid (PLA, Prusament PLA) by fused deposition modeling (FDM) on an Original Prusa i3 MK3S+ (Prusa Research) printer. To ensure a constant gauge length for all samples, the assembly process was performed within a custom 3D-printed alignment jig, after which the cured specimens were transferred to a separate protective fixture, as illustrated in the Supporting Information (Fig. S1). This fixture was subsequently mounted into the machine grips, shielding the fragile wire from torsional and bending forces during the clamping process.

**Table 1**
Chemical composition of alloys used for wire production.

| Element (wt. %) | Zn | Al | Si | Cu | Mn | Fe | Ni | Other |
|---|---|---|---|---|---|---|---|---|
| Mg-0.4Zn | 0.390 | 0.001 | 0.011 | 0.001 | 0.042 | 0.002 | <0.001 | <0.010 |
| Mg-0.6Zn | 0.600 | 0.004 | 0.011 | 0.001 | 0.042 | 0.002 | <0.001 | <0.010 |
| Mg-0.8Zn | 0.800 | 0.003 | 0.011 | 0.001 | 0.042 | 0.002 | <0.001 | <0.010 |
| Mg-1.5Zn | 1.490 | 0.003 | 0.012 | 0.001 | 0.040 | 0.003 | <0.001 | <0.010 |

All specimens were tested at a gauge length of 10 mm and a constant crosshead displacement rate corresponding to a strain rate of $10^{-3}$ $s^{-1}$ using an ElectroPuls E3000 (Instron) electrodynamic testing machine equipped with a 5 kN load cell. Engineering stress and strain were evaluated from the recorded crosshead displacement.

### 2.4. Potentiodynamic polarization

Potentiodynamic polarization (PDP) measurements were performed on extruded Mg-0.4Zn, Mg-0.6Zn, Mg-0.8Zn, and Mg-1.5Zn wires of approximately diameter 300 μm using a Gamry Reference 600 potentiostat. Prior to testing, the wires were pickled in 7 % Nital for 2 min and rinsed in ethanol to remove the surface oxide layer. PDP measurements were performed at 37 °C using a three-electrode setup, with an Ag/AgCl reference electrode (connected via an agar-based salt bridge) and a graphite auxiliary electrode. A 10 cm long wire specimen was used as the working electrode, with a defined exposed surface area. The measurements were carried out in 100 mL of simulated body fluid (SBF), prepared according to Table 2 and adjusted to pH 7.85 at room temperature (approx. 7.4–7.5 at 37 °C, owing to the temperature dependence of Tris). The measurement protocol consisted of a stabilization period of 1200 s, followed by a cathodic polarization scan from +0.05 V to −1.5 V vs. OCP at a scan rate of 1 mV/s. After cathodic polarization, the specimen was allowed to re-equilibrate at OCP for a further 1200 s, after which an anodic polarization scan was performed from −0.05 V to +1.5 V vs. OCP at the same scan rate of 1 mV/s. All scans were recorded with a sampling period of 1 s. Three measurements were performed for each alloy to obtain the mean value.

### 2.5. Degradation in simulated body media

Mg-0.6Zn wires were pickled in a 7 wt% nital solution, as described in Section 2.3, and subsequently immersed in SBF at 37 °C under a constant surface-area-to-volume ratio of 0.3 mL/mm$^2$ for 7 days. Wires were thereafter analyzed using correlative microscopy. First, the products were analyzed via ultraviolet (UV) light source and ET GFP (green fluorescent protein) filter using a Leica M205 FCA fluorescence stereo microscope and Leica DM750 binocular microscope, without any addition of GFP to the medium. Furthermore, EDS was used to obtain the chemical composition of the corrosion products using 5 kV as the accelerating voltage in SEM. The wire was thereafter sputter-coated with gold to improve conductivity for imaging using back-scattered ele ctrons (BSE) in SEM. Micro-Raman spectrometry was used via LabRAM HR Evolution (Horiba) equipped with Nd:YAG excitation laser (532 nm, 100 mW) and air-cooled CCD came ra (Open Electrode, −60 °C) as a detector. Light was forwarded through a 50 × LMPLFLN objective (Olympus, NA = 0.5), using 1800 lines/mm blazed diffraction grating. Spectra were collected in the spectral range from 70 to 2000 cm$^{-1}$, with an acquisition time of 10 – 30 s and 2 accumulations. A few measurements were collected with the maximum range of 4000 cm$^{-1}$, with no relevant peaks detected. To ensure sufficient signal response, the laser exposure of 50 mW was used for fiber-like corrosion products and 100 mW for the rest of the corrosion products. After each measurement, the sample was checked to avoid damage. To ensure reproducibility, different positions on the sample corresponding to individual corrosion products were measured. The measured spectra were truncated to the range 200–1400 cm$^{-1}$ and processed (baseline-correction and smoothing) using asymmetric least squares smoothing (ASLS) coupled with a Savitzky-Golay filter. All depicted Raman spectra were normalized to the maximal intensity peak and collected on the Au-coated sample. In our previous study, we measured the spectra before and after coating with no observed differences in the spectra [26].

X-ray diffraction (XRD) could not be used on wires after immersion due to insufficient signal during the measurement. Therefore, Mg-0.6Zn cylindrical bulk samples, originating from the same ingot as the wires, with a diameter of 14 mm and a height of 3 mm were sectioned from extruded rods with a mean grain size of 19 ± 12 μm. These were ground using a 4000 grit silicon carbide paper, pickled in a 7 wt% nital solution, as described in Section 2.3, and subsequently immersed in SBF (composition given in Section 2.4) at 37 °C for 1, 3 and 7 days. Bulk samples were thereafter analyzed using XRD via powder diffractometer Empyrean (PANalytical) equipped with a cobalt anode as X-ray source. XRD diffractograms were acquired in the range of 10°–90° using grazing incidence diffraction (GID) with reduced penetration depth of the X-radiation, minimizing bulk contribution and enhancing surface sensitivity.

The influence of corrosive environments on the tensile behavior of Mg-0.6Zn wires was assessed using a standardized immersion protocol. Wire segments (35 mm in length) were pickled in a 7 wt% nital solution, as described in Section 2.3, and subsequently immersed at 37 °C under a constant surface-area-to-volume ratio of 0.3 mL/mm$^2$ for 1, 3, and 7 days. All immersion tests were conducted under static conditions, without medium exchange, circulation, or agitation for the entire immersion period. After removal from the medium, the specimens were tensile-tested under the conditions described in Section 2.3. The primary immersion medium was SBF. A subset of Mg-0.6Zn specimens was additionally exposed to a more complex medium consisting of ROTI®-Cell Dulbecco's Modified Eagle Medium (DMEM; Carl Roth GmbH) supplemented with 10 vol% fetal bovine serum (FBS) and 1 vol% penicillin/streptomycin (Capricorn Scientific GmbH), hereafter referred to as DMEM + FBS, and maintained at 37 °C under a 5 % $CO_2$ atmosphere. The pH of the medium was measured at the end of each immersion interval upon specimen removal using an HI6000 (Hanna Instruments) benchtop meter. The fracture surfaces of the tensile-tested specimens were subsequently sputter-coated with gold and examined by JSM-IT500HR SEM (JEOL) using the secondary electron (SE) signal at an accelerating voltage of 20 kV.

**Table 2**
Composition of SBF (according to Müller and Müller [25]) used for PDP.

| Component | Concentration ($g \cdot L^{-1}$) |
|---|---|
| KCl | 0.298 |
| NaCl | 5.840 |
| $NaHCO_3$ | 2.270 |
| $MgSO_4 \cdot 7\ H_2O$ | 0.246 |
| $CaCl_2$ | 0.278 |
| TRIS | 6.060 |
| $KH_2PO_4$ | 0.136 |

## 3. Results

### 3.1. Microstructure and mechanical properties

The quality of the extruded wires was evaluated on the representative Mg-0.6Zn wire to identify factors influencing the mechanical properties and degradation behavior. The presence of oxides was confirmed by SEM and μCT analysis (Fig. 1), with an area fraction of 0.2 % and a volume fraction of 0.05 % (Fig. 1b and e) while the majority of particles exhibited mean size below 2 μm (Fig. 1f). These oxides most likely originated from the alloy itself; however, their exact volume content could not be determined due to insufficient resolution for bulk-scale analysis. The μCT value is therefore a lower estimate (particles close to the 470 nm voxel size are not resolved), whereas the local SEM-SE value tends to be overestimated by the edge and charging contrast of the oxides. In addition to oxides, minor content of other elements was detected in the material microstructure using EDS (Fig. 1e) among which Fe, Mn, and S were the most prevalent. Moreover, a limited number of artifacts artificially increasing the measured oxide volume fraction, were also observed in the wire; these were identified by their characteristic star-like morphology (Fig. 1b, * inset). The wire cross-section was not perfectly circular (Fig. 1c). Grooves formed during extrusion increased the surface roughness, resulting in approximately 4.7 % increase in the actual surface area compared with the theoretical value.

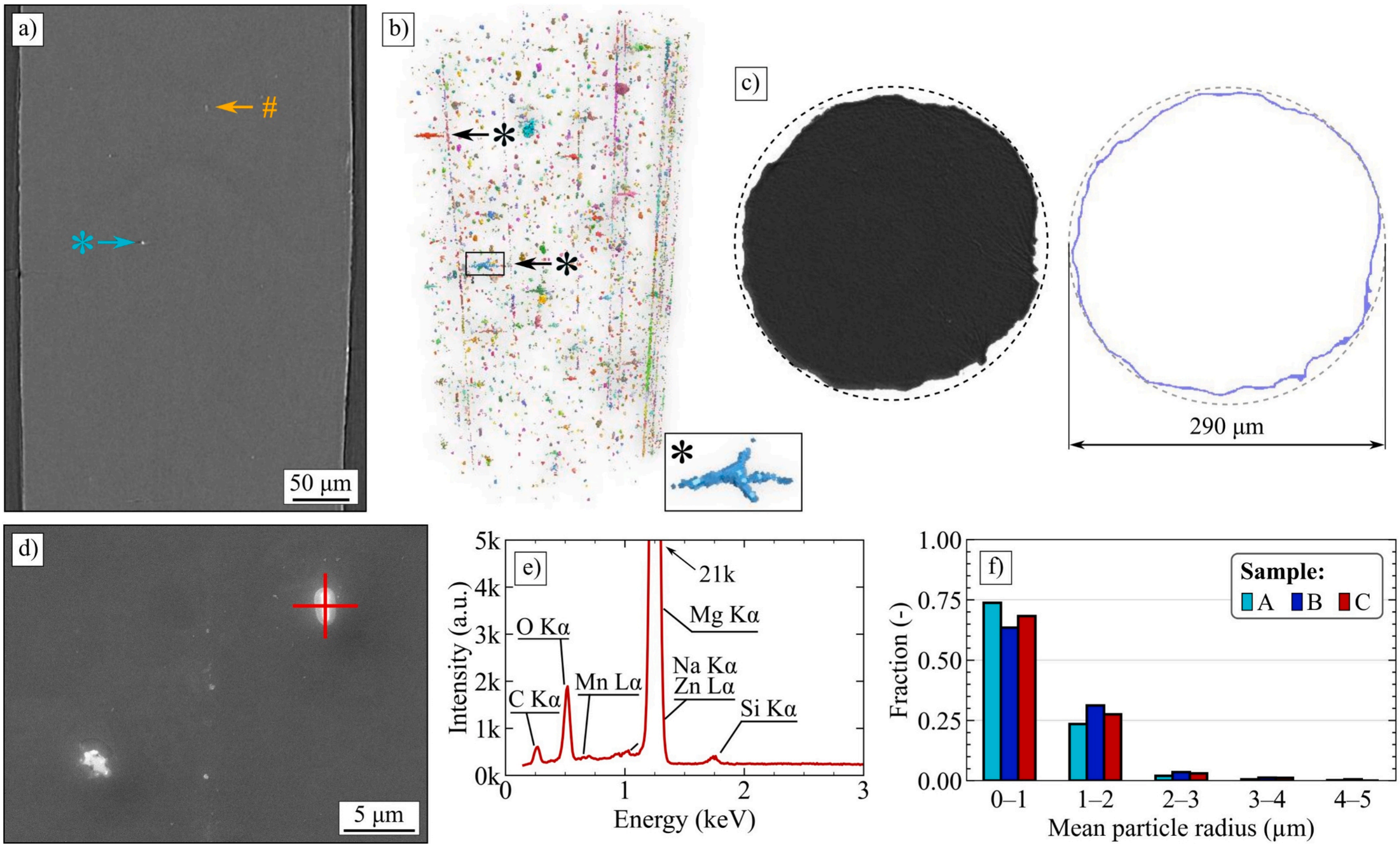

**Fig. 1.** Defects and surface features in extruded wires produced from a commercially prepared alloy. a) 2D μCT image of the wire longitudinal section showing impurities (#) and star-like segmentation artifacts (*). b) μCT reconstruction with highlighted impurity-related features; inset shows a star-like artifact. c) Outline of the actual wire cross-section indicating the increase in surface area compared with the ideal circular geometry. d, e) SEM image and EDS spectrum of oxide particles in the microstructure of the material. f) Impurity particle size distribution determined by μCT.

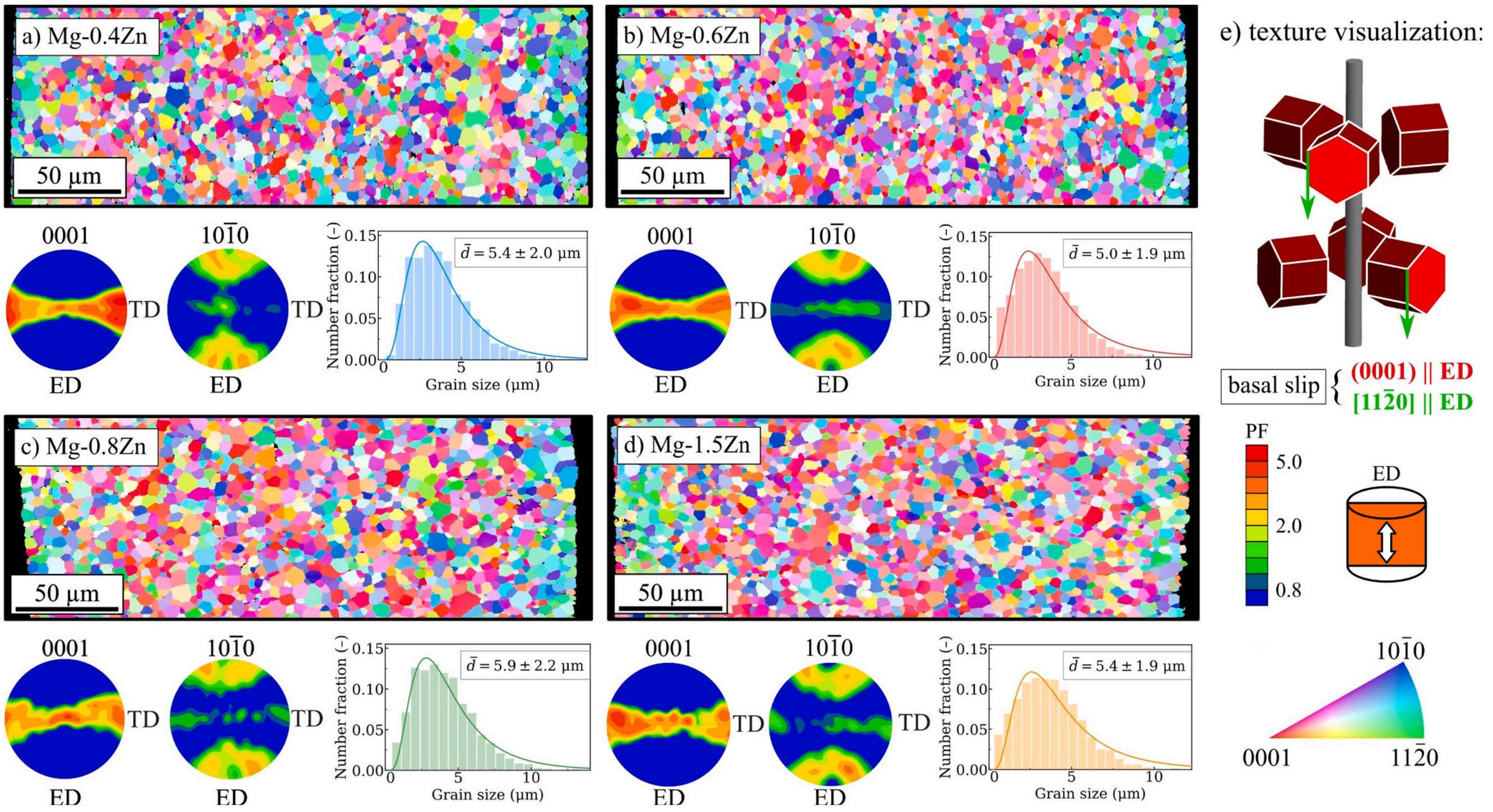

**Fig. 2.** a)–d) IPF map, PF, and histogram of the number fraction of the equivalent circle diameter, including the average grain diameter, for Mg-xZn (x = 0.4, 0.6, 0.8 and 1.5) wire, respectively. e) Illustration of the texture in as-extruded wires.

Inverse pole figure (IPF) maps from longitudinal sections of Mg-xZn (x = 0.4, 0.6, 0.8, 1.5) wires are shown in Fig. 2a–d. The microstructure consists, for all four compositions, of completely recrystallized and predominantly equiaxed grains with a comparable mean grain size of 5.0–5.9 μm (Table 3). The grain size distributions are unimodal and well described by the log-normal fits overlaid on the histograms. The {0001} and $\{10\bar{1}0\}$ pole figures (PF) reveal a moderate basal texture with the c-axis of the grains oriented predominantly perpendicular to the extrusion direction (ED), with the basal planes and two out of six $\{10\bar{1}0\}$ planes aligned parallel to the wire axis, as illustrated in Fig. 2e. The maximum PF densities, expressed in multiples of a random distribution (m.r.d.), reached 5.5 for Mg-0.4Zn, 5.0 for Mg-0.6Zn, 4.2 for Mg-0.8Zn, and 5.0 for Mg-1.5Zn, indicating that the strongest texture was observed in the Mg-0.4Zn wire. Such a texture is commonly observed in extruded Mg alloys [11,27,28].

Schmid factors calculated from the same datasets are nearly identical for all four compositions. For tension along the extrusion direction, the mean values are 0.45–0.46 for prismatic $\langle a \rangle$ slip, and 0.46 for pyramidal $\langle c+a \rangle$ slip, while the $\{10\bar{1}2\}$ extension twin system reaches 0.40–0.41 under compression along the extrusion direction compared with 0.36–0.37 under tension. For basal $\langle a \rangle$ slip the values are 0.158, 0.154, 0.181 and 0.173 for Mg-0.4Zn, Mg-0.6Zn, Mg-0.8Zn and Mg-1.5Zn, respectively, which groups the alloys in the same way as the 0.2 % proof stress (Table 3). In all cases, however, the differences between compositions are smaller than the standard deviation within the individual datasets, which is up to ± 0.12.

As summarized in Table 3, all investigated alloys exhibited comparable tensile properties on nital-pickled specimens ($n \geq 5$ per condition). The 0.2 % proof stress was highest for Mg-0.6Zn (233 ± 6 MPa) and lowest for Mg-1.5Zn (200 ± 7 MPa), whereas the ultimate tensile strength (246–256 MPa) and elongation to failure (23–28 %) were comparable across all compositions and showed no monotonic dependence on Zn content, as depicted in Fig. 3a. A pronounced stress drop immediately after yielding was observed for Mg-0.4Zn and Mg-0.6Zn, while this feature was less distinct in Mg-0.8Zn and Mg-1.5Zn (Fig. 3b).

To study the effect of microstructure and texture on the bending behavior of the wires, Mg-0.6Zn and Mg-1.5Zn wires were selected for observation, as they represent the two ends of the compositional range in which the yield behavior differed most, with a pronounced post-yield stress drop in Mg-0.6Zn and a less distinct one in Mg-1.5Zn (see Table 3 and Fig. 3b). Fig. 4 shows grain orientation spread (GOS) maps, crystal direction (CD) maps, and pole figures (PF) with calculated maximum intensities of the Mg-0.6Zn wire in the as-extruded (AE) state, and after different modes of deformation (for respective inverse pole figure (IPF) maps and kernel average misorientation (KAM) maps see Section 2 in the Supporting Information).

The GOS map for AE wire in the first row of Fig. 4a shows some locally stored residual deformation after direct extrusion. The GOS map was used for qualitative characterization of the Mg-Zn wires post-deformation, providing information on the average misorientation within a given grain. Higher GOS values (red color, see color legend in

**Table 3**
Mechanical properties of extruded nital-pickled Mg-Zn wires determined by uniaxial tensile testing. $\bar{d}$– average grain size; $\sigma_{02}$ – 0.2 % proof stress; $\sigma_m$– ultimate tensile strength; $A_t$– total elongation to failure. Values are reported as mean ± standard deviation.

| Alloy | $\bar{d}$(μm) | $\sigma_{02}$ (MPa) | $\sigma_m$ (MPa) | $A_t$ (%) |
|---|---|---|---|---|
| Mg-0.4Zn | 5.4 ± 2.0 | 229 ± 7 | 249 ± 4 | 23 ± 5 |
| Mg-0.6Zn | 5.0 ± 1.9 | 233 ± 6 | 252 ± 3 | 27 ± 2 |
| Mg-0.8Zn | 5.9 ± 2.2 | 207 ± 6 | 246 ± 4 | 28 ± 3 |
| Mg-1.5Zn | 5.4 ± 1.9 | 200 ± 7 | 256 ± 6 | 25 ± 4 |

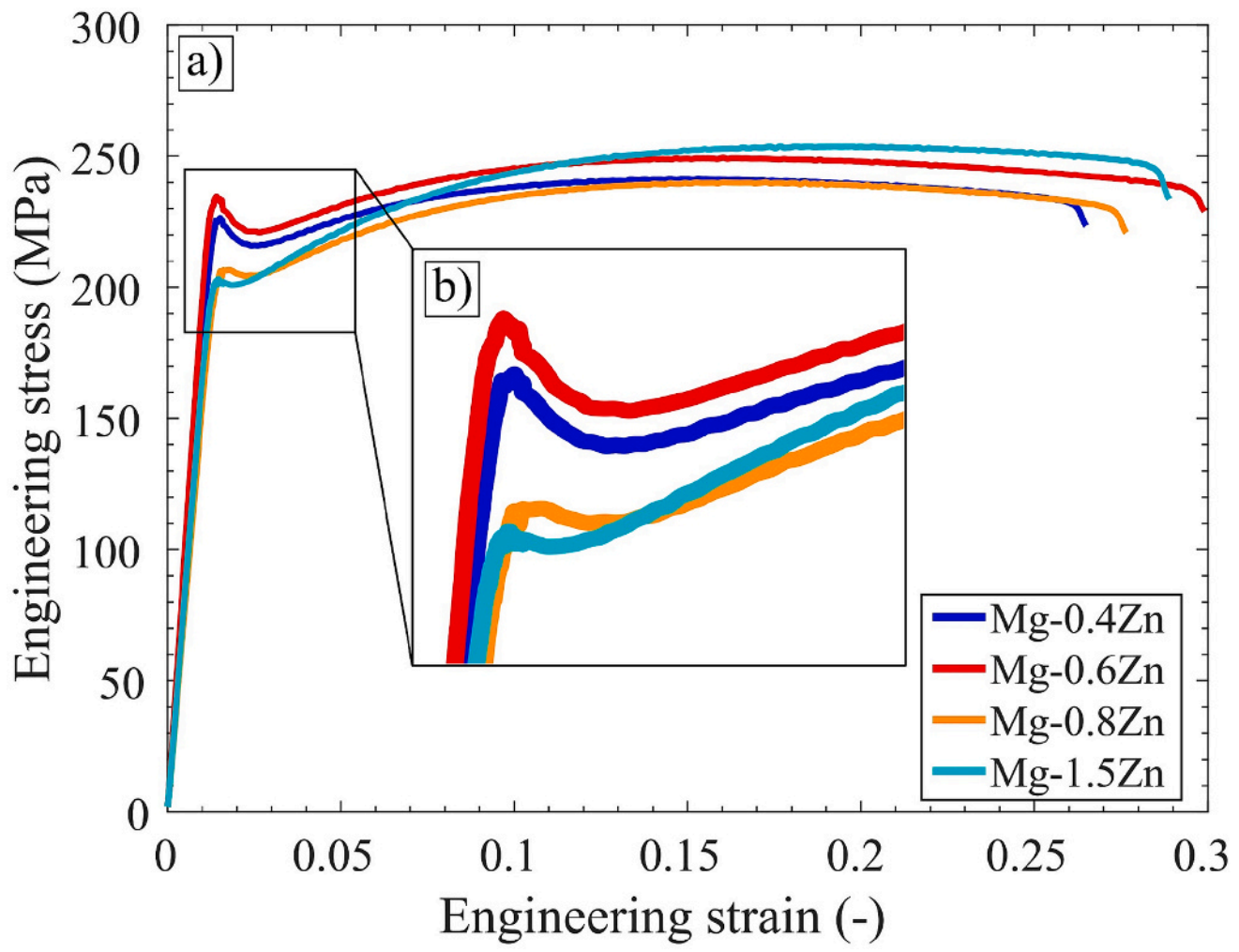

**Fig. 3.** Representative engineering stress–strain curves of extruded nital-pickled Mg-xZn (x = 0.4, 0.6, 0.8 and 1.5) wires: a) full curves; b) magnified view of the elastic–plastic transition, showing a pronounced post-yield stress drop in Mg-0.4Zn and Mg-0.6Zn and a less distinct one in Mg-0.8Zn and Mg-1.5Zn.

Fig. 4) indicate generally higher accumulated strain or the presence of partially twinned grains. For further discussion of the GOS values, see Section 4.2. CD map and PF for {0001} and $\{10\bar{1}0\}$ in the second and third row of Fig. 4a showcase a strong texture discussed before. The CD map of $\langle 0001 \rangle$ represents c-axis orientation perpendicular to the extrusion direction as dark red (see color legend in Fig. 4). Bright green represents c-axis oriented parallel to the ED, which in the case of Mg hexagonal close-packed (hcp) matrix approximately corresponds to $\{10\bar{1}2\}$ tensile twins due to the misorientation angle being 86° relative to the matrix.

Uniaxial tensile stress during the tensile test (TT) along the ED leads to a uniform increase in GOS in the wire cross-section after 2 % and 20 % deformation (Fig. 4b and c). Furthermore, the maximum PF intensity increases from 5.0 m.r.d. (Fig. 2b) or 6.1 m.r.d. (Fig. 4a) in the AE wire to 9.2 m.r.d. after 20 % deformation (Fig. 4c). {0001} PF sharpens with increasing deformation, while $\{10\bar{1}0\}$ PF showcases a decrease in texture compared to the AE wire.

The deformation behavior of the Mg-0.6Zn wire after bending can be seen in Fig. 4e–h, with the custom bending apparatus and EBSD acquisition regions being shown in Fig. 4d. Bending of the wire leads to asymmetric plastic deformation relative to the wire centerline, with the neutral zone, exhibiting minimal deformation, being shifted away from the center of curvature in Fig. 4e towards the tension zone (TZ). Moreover, the compression zone (CZ) of the wire shows severe tensile twinning based on the CD map of the wire, while the TZ shows no change in crystal direction compared to the AE wire or the wire after TT. Following bending, the PF intensity decreases due to the formation of twins, reorienting the c-axis approximately parallel to the extrusion direction, as seen in the {0001} PF, while the $\{10\bar{1}0\}$ PF remains qualitatively intact.

Straightening of the wire after bending is shown in Fig. 4f. Relaxation of stress around the centerline with widening of the neutral zone is visible in the GOS map, retaining a slight asymmetry of the neutral zone placement in the direction away from the previous center of curvature. The CD map shows secondary tensile twinning of the twins formed in the previous compression zone during the bending, redirecting the crystal structure to the AE state with {0001} basal planes being oriented parallel to the ED. This so-called twinning-detwinning mechanism allows for reversible plasticity during bending of the Mg-Zn wires and is in line with the findings on pure Mg [11]. Tensile twins form again in the new

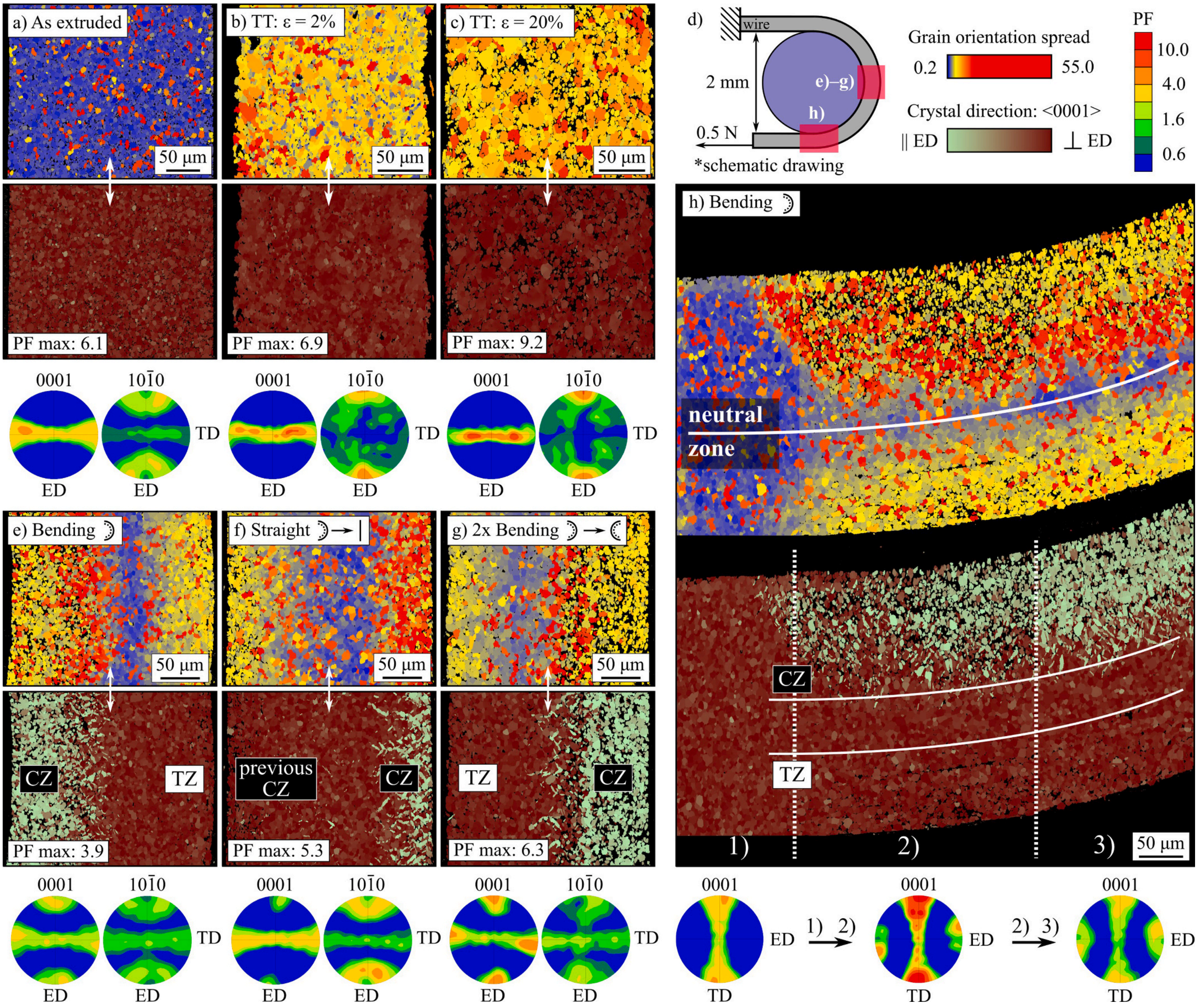

**Fig. 4.** EBSD analysis of Mg-0.6Zn wire deformation showing GOS maps, CD maps of ⟨0001⟩, and PFs (extrusion direction – ED; transversal direction – TD), with their corresponding maxima shown as an inset in the CD maps, for: (a) as-extruded wire; (b) and (c) wire after 2 % and 20 % deformation (ε) during tensile test (TT), respectively; (e)–(g) wire after bending, subsequent straightening, and bending in the opposite direction, respectively (TZ and CZ denote the tensile and compression zone). (d) Schematic drawing of the bending apparatus using a diameter of 2 mm, where red rectangles denote acquisition areas for (e)–(h). (h) GOS map, CD map of ⟨0001⟩, and PFs after single bending (PFs were calculated from areas indicated by numbers 1–3). (For interpretation of the references to colour in this figure legend, the reader is referred to the web version of this article.)

CZ, but to a much lesser extent compared to the bending mode shown in Fig. 4e due to lower deformation energy.

Bending of the wire with subsequent bending in the opposite direction is depicted in Fig. 4g, showing nearly identical qualitative behavior in GOS and CD maps compared to the first bending, with shifting of the neutral zone away from the centerline and the formation of tensile twins on the compressive side. However, the PF maximum intensity is higher after the second bending, with a slight decrease in the {10$\bar{1}$0} PF texture.

The transition zone of the wire from the non-deformed region to the bending-affected region is shown in Fig. 4h. The map was collected and merged from three separate EBSD measurements, denoted as 1)–3), which are separated by the white dotted lines. Corresponding PFs were determined from these three regions. The non-deformed region shows GOS and CD maps similar to the as-extruded wire, with a slight GOS increase on both wire surfaces due to the experimental setup. Within zone 2) with a length of approximately 300 μm in the ED direction, saturation of GOS and the tensile twin density can be seen. From zone 3) onwards, GOS and CD maps become similar to Fig. 4e and g. The slight increase in the GOS map intensity in the neutral zone (less intense blue color, both in Fig. 4g and Fig. 4h) is caused primarily by the quality of the signal during EBSD (for detailed discussion, see Section 2 in the Supporting Information).

The Mg-1.5Zn wire was subjected to the same bending modes as the presented Mg-0.6Zn wire. No qualitative change in deformation behavior was observed after bending, straightening, and bending to the opposite direction (see Section 2 in the Supporting Information , Fig. S3), which is explainable by the near-identical microstructure depicted in Fig. 2. The slight increase of 33 MPa in the 0.2 % proof stress of the Mg-0.6Zn wire compared to the Mg-1.5Zn wire was insufficient to alter the bending behavior observed by EBSD.

### 3.2. Corrosion behavior

Semilogarithmic plots of the PDP data for one series of wires are presented in Fig. 5. Corrosion potential $E_{corr}$ was determined from the stabilized open circuit potential prior to polarization. Due to the high reactivity of Mg-based alloys in SBF, a shift in OCP was observed between the measurements of the cathodic and anodic branches, indicating a non-stationary system.

The corrosion current density $i_{corr}$ was evaluated using Tafel extrapolation. Because of the OCP shift and the progressive surface modification during polarization, the cathodic branch was used for the extrapolation, as it is less affected by prior polarization and is considered to better approximate the corrosion state. Furthermore, the anodic branch of magnesium is known to be affected by the negative difference effect (NDE), where the measured current represents a superposition of anodic dissolution and an increasing cathodic hydrogen evolution reaction. As a result, the anodic branch does not reflect purely anodic kinetics and is not suitable for reliable Tafel analysis.

Linear regions were identified on cathodic branches within a polarization range of 50–150 mV from the OCP. The segment with the highest coefficient of determination ($R^2$) within this range was used for extrapolation. The resulting corrosion current density $i_{corr}$, cathodic Tafel slope $b_c$, and linear corrosion rate CR in mm/y are reported in Table 4. In addition, polarization resistance $R_p$ was evaluated within the polarization range of ± 10 mV from the OCP.

It should be noted that the calculated corrosion rate assumes uniform material loss and linear degradation over time. These assumptions are not strictly valid for magnesium due to its localized and time-dependent corrosion behavior. Therefore, the reported corrosion rates should be considered comparative rather than absolute values.

Since the polarization measurements revealed no pronounced differences between the four compositions, the selection of a single alloy for the detailed characterization of degradation behavior and of the associated loss of mechanical integrity was based on the tensile results. Although all four alloys performed comparably, Mg-0.6Zn offered the most favorable overall combination of 0.2 % proof stress, ultimate tensile strength, and elongation to failure, reaching the highest proof stress of the series while retaining strength and ductility at the upper end of the measured range (Table 3), and was therefore chosen.

Localized regions with extensive corrosion product formation are analyzed using correlative microscopy in Fig. 6a–c. The SEM-BSE image in Fig. 6a is slightly magnified and rotated along the wire axis relative to images b and c due to sample manipulation between image acquisitions. Large corrosion products overgrown with fiber-like products were observed (see inset in Fig. 6a). Both corrosion products emit autofluorescence under UV light, while the part of the wire with a thin blue layer in Fig. 6b emits no fluorescence (black color on the wire in Fig. 6c).

**Table 4**
Results of PDP evaluated from cathodic branches measured for extruded wires in SBF at 37 °C. $E_{corr}$ – corrosion potential, $b_c$ – cathodic Tafel slope, $R_p$ – polarization resistance, $i_{corr}$ – corrosion current density, CR – corrosion rate.

| Alloy | $E_{corr}$ (V) | $b_c$ (mV/dec) | $R_p$ ($\Omega\cdot cm^2$) | $i_{corr}$ ($\mu A/cm^2$) | CR (mm/y) |
|---|---|---|---|---|---|
| Mg-0.4Zn | −1.85 ± 0.05 | 320 ± 60 | 207 ± 7 | 194 ± 9 | 4.5 ± 0.2 |
| Mg-0.6Zn | −1.82 ± 0.01 | 240 ± 20 | 160 ± 20 | 220 ± 30 | 5.1 ± 0.6 |
| Mg-0.8Zn | −1.78 ± 0.05 | 330 ± 40 | 192 ± 18 | 233 ± 15 | 5.3 ± 0.4 |
| Mg-1.5Zn | −1.81 ± 0.02 | 310 ± 50 | 179 ± 10 | 230 ± 30 | 5.5 ± 0.7 |

Certain areas of the wire in Fig. 6d show the formation of a corrosion layer with cracking, which is intertwined with localized regions exhibiting a different corrosion mechanism (denoted by the dark blue arrows). The depicted cracking and delamination of the corrosion layer in SEM imaging could be caused by the high vacuum during acquisition. However, severe cracking was observed before SEM imaging, as shown in Fig. 6h, and could have formed during the immersion itself, or might be an artefact of sample drying after immersion. More detailed in-situ observation of the wire during and after the immersion will be carried out in future work to determine the onset of cracking in this layer. In addition to the localized variations in the corrosion mechanism, some parts of the wire featuring the corrosion layer were completely covered by fibrous corrosion products, as depicted in Fig. 6e with a detail of the fibers shown in Fig. 6f. Other areas formed a uniform layer of corrosion products with observed autofluorescence in the cracks of this layer (Fig. 6g and h).

The Raman and EDS spectra of corrosion products are depicted in Fig. 6i and j. Corrosion products marked by blue and red crosses have nearly identical Raman spectra in the observed range. The peaks at 430 $cm^{-1}$, 580 $cm^{-1}$, and 950 $cm^{-1}$ are characteristic of hydroxyapatite (HA) according to Nelson and Williamson [29] and the RRUFF database (R060180). The most intense peak at 950 $cm^{-1}$ corresponds to the totally symmetric stretching mode ($\nu_1$) of the $PO_4$ group, the 430 $cm^{-1}$ peak to the tetrahedra's $PO_4$ group doubly degenerate bending mode ($\nu_2$), and the 580 $cm^{-1}$ peak to the triply degenerate bending mode of the $PO_4$ group [30]. The corrosion product denoted by the symbol # has a Ca:P ratio of 1.60, nearly identical to the ratio 1.67 reported for HA by Koutsopoulos [31], confirming the presence of HA.

The fibrous corrosion products show a strong peak at 950 $cm^{-1}$, characteristic of HA. However, there is a major asymmetry of the peak toward higher Raman shifts, denoted by the light-blue arrow in Fig. 6i.

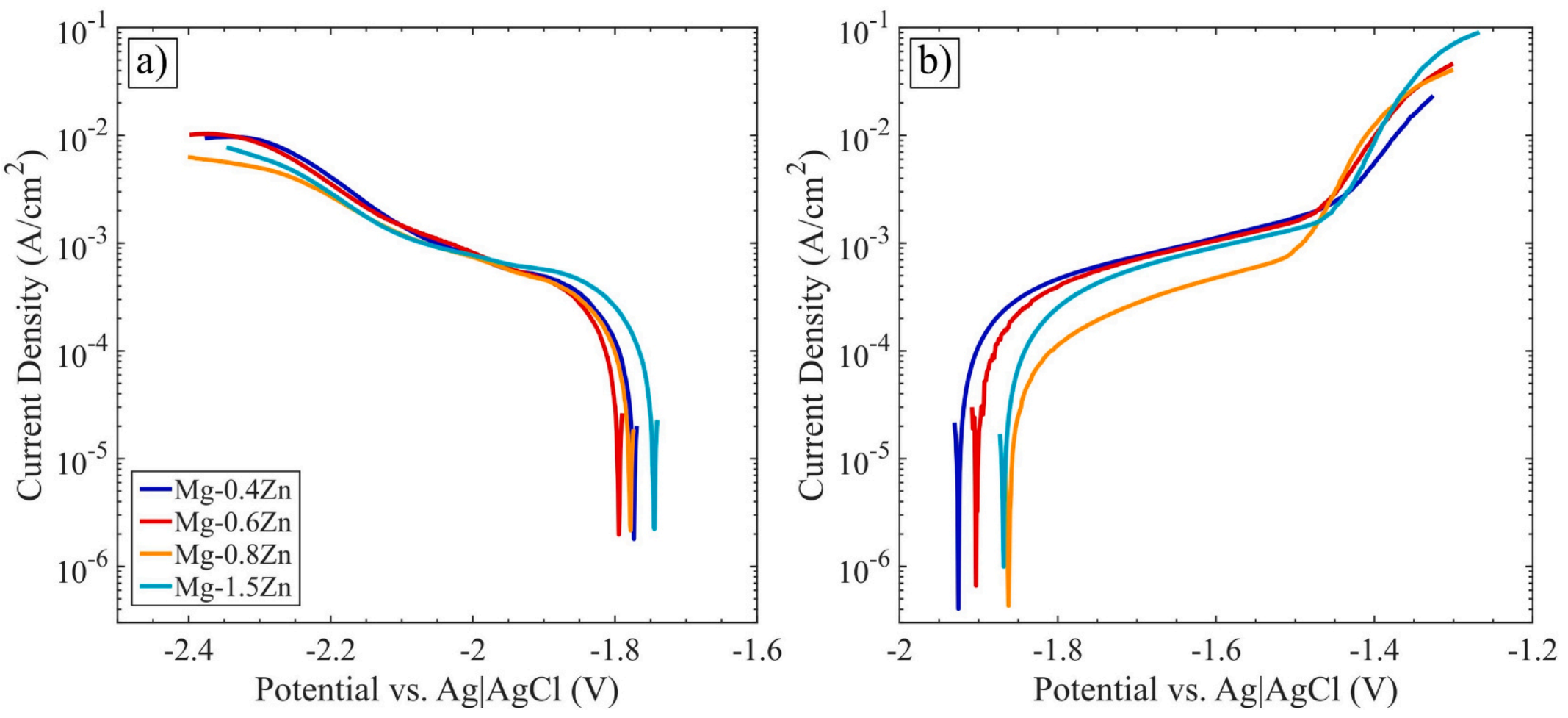

**Fig. 5.** Representative potentiodynamic polarization curves of Mg-xZn (x = 0.4, 0.6, 0.8 and 1.5) wires in SBF: (a) cathodic branches; (b) anodic branches.

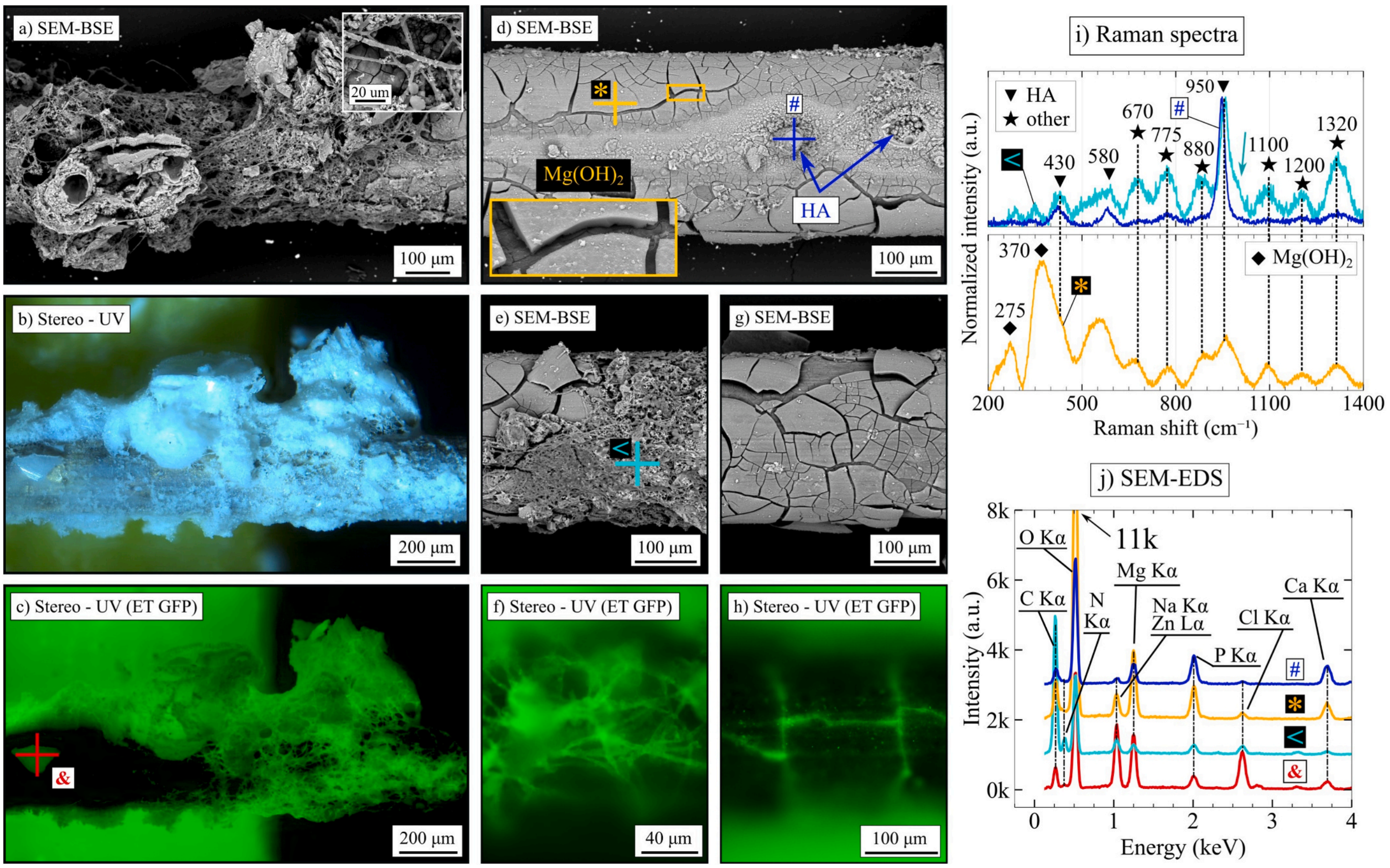

**Fig. 6.** (a)–(h) Analysis of corrosion products formed on Mg-0.6Zn wire after degradation in SBF for 7 days using SEM-BSE and a stereo microscope (for details, see Section 2.5). (i), (j) Raman and EDS spectra from images (c), (d), and (e). Each spectrum is denoted by color and a symbol, which corresponds to the corrosion product depicted in the SEM-BSE or stereomicroscope image.

Moreover, EDS revealed a much higher amount of C and a very small amount of Ca, ruling out the presence of HA. Although the measurements of carbon via EDS are often misleading, combined with Raman these findings suggest the presence of B-type carbonated apatite, in which carbonate $CO_3^{2-}$ substitutes for phosphate ions, leading to lower Ca concentrations. McElderry et al. [32] observed a broadening of the peak for the mode ($\nu_1$) of the $PO_4^{3-}$ group, which was more pronounced with increasing $CO_3^{2-}$ content. The 670 cm$^{-1}$ peak in Fig. 6i is also likely attributable to carbonated apatite as the in-plane bending vibration ($\nu_4$) of the $CO_3^{2-}$ [29].

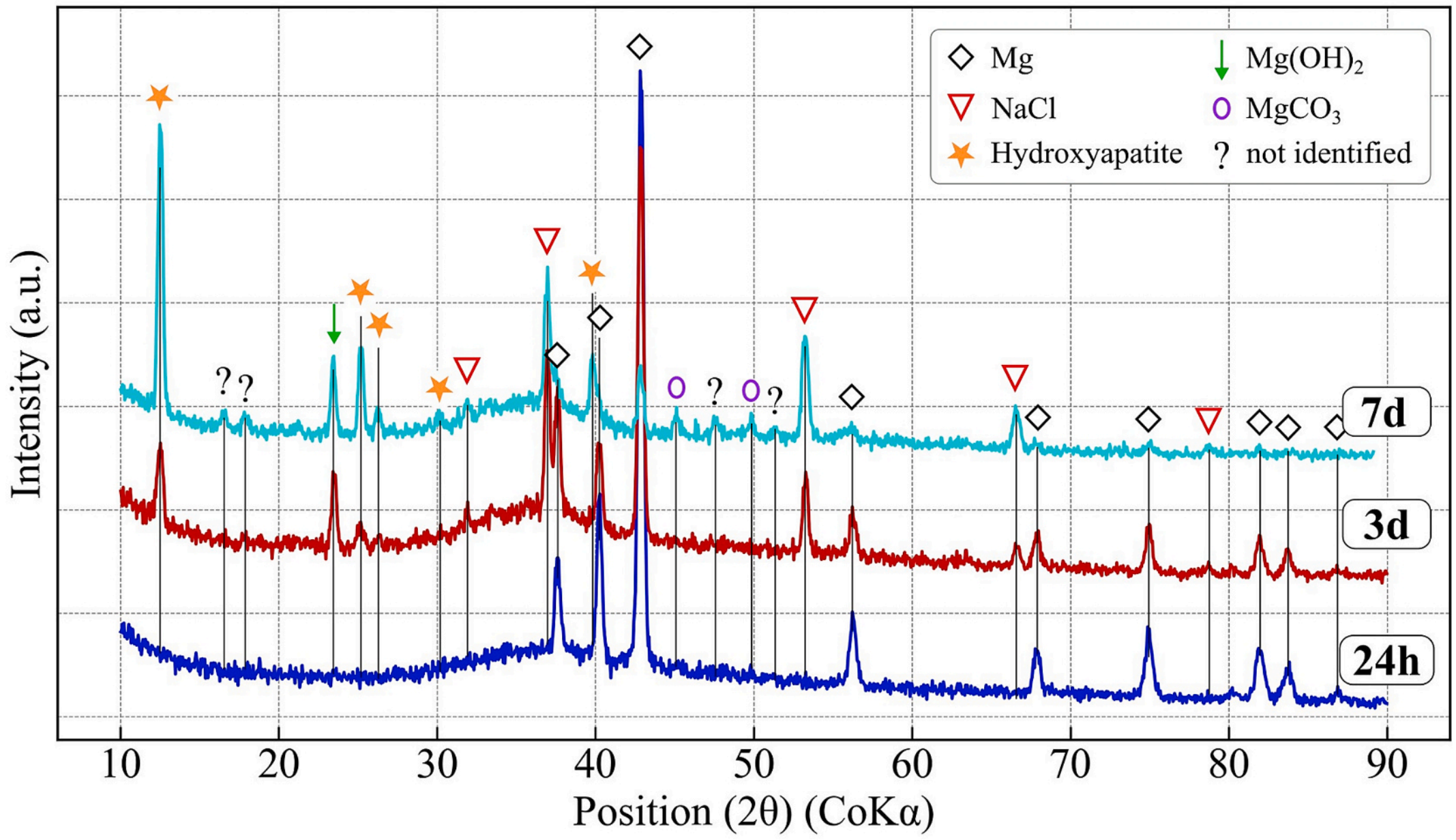

**Fig. 7.** X-ray diffraction patterns of bulk Mg-0.6Zn samples after immersion in SBF for 24 h, 3 d, and 7 d.

Several less intense peaks were observed in the spectrum of the fibrous product. The 1320 $cm^{-1}$ and 1200 $cm^{-1}$ peaks are likely remnants from the SBF immersion [33,34], with the 1100 $cm^{-1}$ peak attributed to the P-O-C stretching [35]. The P-O-P bond and its asymmetric stretching vibration mode ($\nu_{as}$) is present as a 775 $cm^{-1}$ peak [36].

The spectrum of the corrosion layer (Fig. 6i) suggests the presence of a brucite ($Mg(OH)_2$) layer, with a mixed OH stretch peak at 370 $cm^{-1}$, and the 275 $cm^{-1}$ peak corresponding to an $E_g$ lattice vibrational mode associated with translational motions of brucite layers [37,38]. However, the most intense peak for $Mg(OH)_2$ at around 443 $cm^{-1}$ is not present, and a slight asymmetry can be observed in the 370 $cm^{-1}$ peak, phenomena that might be connected to the presence of brucite in a highly disordered state [37], as expected from a corrosion product.

Fig. 7 shows XRD results measured on bulk Mg-0.6Zn samples after three immersion periods in SBF. After 24 h of immersion, no corrosion products could be identified by XRD; however, the GID scan revealed the presence of an amorphous layer based on the bulge observed in the range of 30°–45°.

XRD confirmed the presence of hydroxyapatite and $Mg(OH)_2$ (observed in Fig. 6) after 3 and 7 days of immersion. Furthermore, a high amount of NaCl was observed, with minor diffraction peaks that might suggest the presence of magnesite ($MgCO_3$). However, neither EDS analysis nor Raman spectroscopy identified such a phase on the wire sample, and the verification of its presence remains to be confirmed. The higher NaCl content on the bulk sample was likely caused by the recrystallization of dissolved salts from the SBF solution during drying. Due to the larger surface area and the nature of the formed products, more of the compound was observed on the bulk sample compared to the wire sample. After longer immersion periods of 3 and 7 days, the volume of corrosion products formed on the bulk samples gradually increased, replacing Mg in the sample, as indicated by the decreasing diffraction intensity of Mg and more pronounced peaks of the corrosion products.

### 3.3. Loss of mechanical properties after degradation

Tensile testing of Mg-0.6Zn wires after immersion in SBF revealed rapid deterioration of mechanical properties. Because corrosion progressed non-uniformly along the wire surface, the load-bearing cross-section could not be reliably quantified, and engineering stress became an unreliable descriptor; mechanical performance is therefore reported as tensile force rather than stress [18]. As shown in Table 5, degradation in SBF was severe. After 7 days of immersion, five of the six immersed Mg-0.6Zn specimens had dissolved completely (the sixth was too degraded to be tested) and the medium pH had risen from an initial 7.85 to 9.5, precluding mechanical testing of this group. Pronounced degradation was already evident after 3 days: $F_M$ and $F_{02}$ dropped by 41 % and 40 %, respectively, while elongation at fracture $A_t$ collapsed from 27 ± 2 % to 1 ± 1 %, a 95 % reduction indicating near-complete loss of ductility. The comparatively large dispersion of the 3-day values is consistent with the stochastic nature of the localized attack evident on the fracture surface (Fig. 9f), in which the position and depth of individual pits govern the residual load-bearing cross-section of each specimen. In DMEM + FBS, by contrast, $F_{02}$ and $F_M$ decreased mainly within the first day and changed little thereafter, while $A_t$ decreased continuously to 9 ± 3 % after 7 days (Table 5).

**Table 5**
Mechanical properties of Mg-0.6Zn wires after immersion in SBF and DMEM + FBS at 37 °C for 1, 3 and 7 days. $A_t$ – total elongation to failure; $F_{02}$ – 0.2 % offset yield force; $F_M$ – maximum tensile force. Values are reported as mean ± standard deviation (n = 3; 0 days: n ≥ 5, see Table 3).

| Immersion time (days) | $F_{02}$ (N) | $F_M$ (N) | $A_t$ (%) |
|---|---|---|---|
| 0 | 14.3 ± 0.4 | 15.5 ± 0.2 | 27 ± 2 |
| SBF | | | |
| 1 | 11.4 ± 1.1 | 13.1 ± 1.5 | 12 ± 4 |
| 3 | 8.6 ± 1.7 | 9.2 ± 2.1 | 1 ± 1 |
| 7 | – | – | – |
| DMEM + FBS | | | |
| 1 | 11.9 ± 0.2 | 14.0 ± 0.3 | 19 ± 4 |
| 3 | 11.8 ± 0.4 | 14.0 ± 0.6 | 14 ± 2 |
| 7 | 11.8 ± 0.3 | 13.4 ± 0.2 | 9 ± 3 |

## 4. Discussion

### 4.1. Effect of wire quality

Microstructural observations confirmed the presence of impurity elements known to influence the degradation behavior of Mg, particularly Fe, S, and Mn. These elements originate from compounds used for the reduction of MgO to Mg and may also be partially incorporated into slag during recycling of Mg alloys. However, complete removal is not achieved, and a fraction of these elements remains in the final alloy. The impact of these elements on degradation behavior is strongly dependent on their concentration and spatial distribution within the microstructure. Based on the nominal composition provided by the manufacturer and the statistical occurrence determined by EDS analysis, their presence is limited to a relatively small number of localized heterogeneities. In contrast, oxide particles were observed to be distributed throughout the entire material volume. Based on this, oxide particles are expected to have the most pronounced effect on both degradation and mechanical behavior due to the lattice mismatch and non-coherent regions at the Mg-MgO interface [39]. Such interfaces can act as stress concentrators because of inefficient stress transfer during mechanical loading and thus decrease overall performance. Moreover, these regions may serve as preferential sites for the initiation of localized corrosion due to the formation of a more complex system with multiple interfaces susceptible to attack. Since all four alloys derive from the same base material and share nearly identical impurity contents (Table 1), the oxide content and distribution are expected to be effectively constant across compositions; any oxide contribution would therefore be common to all alloys and is unlikely to account for the composition-dependent behavior observed here. Despite the anticipated negative effects of oxides, no such effect could be clearly isolated in the present data.

### 4.2. Deformation behavior

Hot-extruded Mg–Zn alloys are commonly characterized by fully recrystallized, equiaxed grain structures free of secondary phases [40]. The extruded Mg-Zn wires exhibit a pronounced crystallographic texture gradient from the surface to the core, as illustrated in Fig. 8 for the Mg-0.6Zn wire, evaluated separately for individual sub-areas of the same longitudinal section as in Fig. 2b; the remaining alloys display analogous behavior. The outer regions (Fig. 8a and b) develop a strongly aligned fiber texture with maximum intensities reaching 19 m.r.d., whereas the core (Fig. 8c) exhibits a substantially weaker basal texture with maximum intensities of only 5.5 m.r.d., a condition representative of the majority of the wire bulk. This gradient arises from the heterogeneous distribution of shear stress imposed during extrusion: the outer regions are subjected to severe frictional shear against the die wall, which forces grains into a strongly aligned and well-ordered fiber texture. In contrast, grains within the core experience reduced shear constraint and retain greater rotational freedom about the extrusion axis, resulting in a more disordered fiber texture characterized by a broader distribution of crystallographic orientations [41].

Peng et al. [42] reported that, in backward extruded Mg–Zn alloys, increasing the Zn content from 0.5 to 2 wt% resulted in progressive grain refinement, with average grain sizes of 47.5, 37.5, 36, and 23 μm for Mg-0.5Zn, Mg-1Zn, Mg-1.5Zn, and Mg-2Zn, respectively. This microstructural refinement was accompanied by a clear increase in tensile strength, with UTS values rising from 145 ± 8 MPa to 169 ± 9 MPa, 190 ± 7 MPa, and 198 ± 6 MPa, respectively. When considered together with the present results, where increasing Zn content produced no significant grain-size variation and only limited changes in strength, it appears that the strengthening response of these dilute Mg–Zn wires is governed primarily by the final grain size rather than by Zn content alone. In other words, once a sufficiently high accumulated equivalent strain is introduced during processing, further improvement in mechanical properties is expected to be controlled mainly by the extent of grain refinement, achieved by limiting grain growth during and after the processing. Therefore, for the present wire-extruded alloys, further optimization should focus on adapting the extrusion process to promote more effective microstructural refinement, or on introducing additional alloying elements through microalloying. This fact allows for the selection of Zn content to be matched mainly to the biological response of these implants rather than the mechanical properties alone.

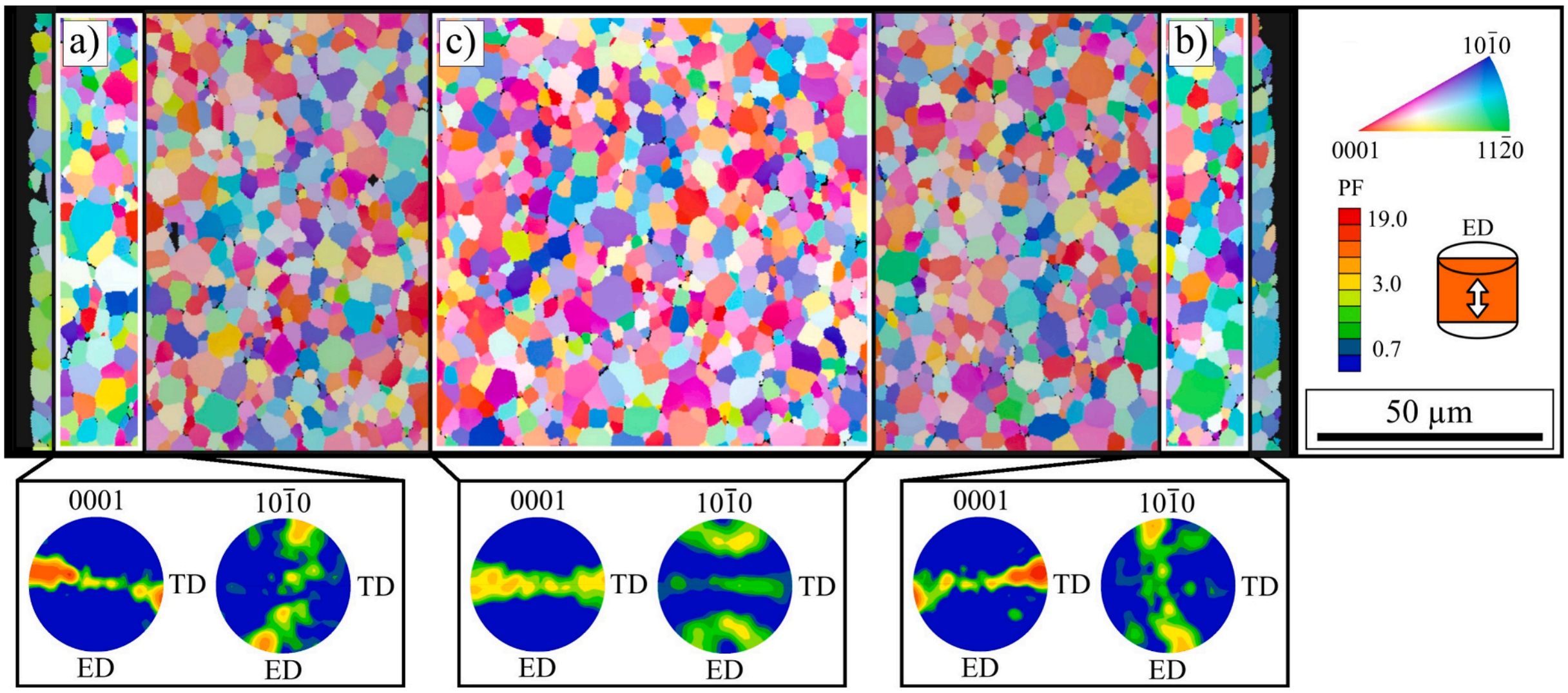

**Fig. 8.** IPF map of the longitudinal section of an as-extruded Mg-0.6Zn wire, with corresponding {0001} and {$10\bar{1}0$} pole figures (the same section as in Fig. 2b): (a) and (b) wire edges, exhibiting a strong fiber texture with maximum density up to 19 m.r.d.; (c) wire core, displaying a weaker basal texture with maximum density up to 5.5 m.r.d.

A plausible contributing factor may be the experimental extrusion configuration employed here, in which die temperature could be regulated only prior to the extrusion step, potentially allowing a thermal offset to develop between alloys during processing. Although this hypothesis remains to be verified experimentally, it suggests that precise and continuous die temperature control may be beneficial for ensuring manufacturing consistency across compositions.

Conversely, higher Zn content enhanced the strain hardening capacity, defined as $H_c = (\sigma_m - \sigma_{02})/\sigma_{02}$ [43], with values of 0.09, 0.08, 0.19 and 0.28 for Mg-0.4Zn, Mg-0.6Zn, Mg-0.8Zn and Mg-1.5Zn, respectively. It should be noted that this increase results from the lower proof stress of Mg-0.8Zn and Mg-1.5Zn rather than from an increasing ultimate tensile strength, which remained constant within scatter (246–256 MPa). The Schmid factors in Section 3.1 are effectively the same for all four alloys and the texture maxima do not decrease monotonically with Zn content, so that the change in $H_c$ is not explained by a systematic weakening of the basal texture or by a composition-dependent availability of non-basal slip. Direct identification of the active slip systems would require in-situ diffraction or TEM analysis, which was not performed here.

The tensile test results indicate that all the studied alloys exhibit a promising combination of tensile strength and total elongation to failure for biomedical applications (Table 6). Notably, the extruded wires investigated in this study achieve a favorable balance between tensile strength and ductility, with elongation to failure exceeding that reported for most cold-drawn wires of comparable Mg-based alloy systems. This enhanced ductility can be attributed to the fine grain size in combination with a moderately strong fiber texture [46]. All alloys exhibited a yield drop, most pronounced for Mg-0.4Zn and Mg-0.6Zn. As described by Drozdenko et al. for Mg–Zn–Y alloys, this phenomenon originates from the sudden release of non-basal dislocations previously pinned by stacking faults and solute atoms [47]. A comparable mechanism, in which dislocation pinning together with a fine grain size governs the onset of the yield drop, has likewise been reported for other Mg-based alloy systems [48,49]. In the present Mg–Zn system, the pinning mechanism is most plausibly governed by Zn segregation, as proposed in our previous work [18], where we did not observe a comparable yield drop in pure Mg wires processed under analogous conditions.

**Table 6**
Comparison of mechanical properties of thin Mg-alloy wires for biomedical applications. Dia. – final wire diameter, E/D – extruded/drawn condition, $\sigma_{02}$ – 0.2 % proof stress, $\sigma_m$ – maximum engineering stress, $A_t$ – total elongation to failure. Approximate values marked with ~ were roughly deduced from the presented graphs.

| Alloy | Dia. (μm) | E/D | $\sigma_{02}$ (MPa) | $\sigma_m$ (MPa) | $A_t$ (%) | Ref. |
|---|---|---|---|---|---|---|
| Mg-0.6Zn | 280 | E | 233 ± 6 | 252 ± 3 | 27 ± 2 | This work |
| Mg | 250 | E | 198 ± 2 | 246 ± 4 | 10 ± 1 | [11] |
| ZXM100 | 300 | D | 260 ± 7 | 284 ± 12 | 18 ± 1 | [15] |
| ZX10 | 400 | E | 154 ± 2 | 217 ± 1 | 34 ± 1 | [12] |
| Mg-6Zn | 310 | D | ~220 | 300 ± 10 | 11 ± 2 | [44] |
| Mg-5Zn | 230 | D | 245 | 303 | 23 | [45] |

It is important to note that the highest grain orientation spread in Fig. 4 (red grains) does not always correspond to the highest deformation in the given grain. High GOS intensity is caused either by a grain with partial twinning or a grain with a high residual strain. For the wires deformed during the tensile test with no observed twinning, higher GOS values are caused by higher accumulated deformation energy, which is confirmed by KAM maps in Section 2 of the Supporting information. In the compression zones of the bent wires with grains fully twinned (c-axis parallel to the extrusion direction), GOS intensity decreases. Apart from the grains with high strains (Fig. 4f, previous CZ), the most intense GOS regions post-bending are located in the areas with partially twinned grains, which can be best seen during twin density increase and saturation in the CZ of Fig. 4h, area 2).

Post-bending, GOS maps revealed the shifting of the neutral zone away from the center of curvature in Fig. 4e–h, with twinning in the compression zone and no twinning in the tensile zone, which is consistent with the well-documented behavior of Mg-alloys during bending [50,51]. Ren et al. [52] observed a dependence of the initial texture on the twinning formation and shifting of the neutral layer. During three-point bending of an Mg-3Al-1Zn alloy with a strong crystallographic texture, loaded perpendicular to the c-axis, the neutral layer shifted toward the compression zone, accompanied by a similar twinning asymmetry as observed in the present study. When the load was applied in the direction parallel to the c-axis for a different sample texture, shifting of the neutral layer occurred in the opposite direction, towards the tensile zone. This asymmetry in the neutral layer shifting is partially explained by the tension–compression asymmetry of Mg alloys [52]. The plastic deformation, occurring dominantly by $\{10\bar{1}2\}$ twinning and prismatic $\langle a\rangle$ slip in Mg-alloys [53], occurs in a larger area in the CZ, compared to the TZ, causing the shifting of the neutral layer.

The asymmetric activation of deformation mechanisms during the bending of Mg alloys is governed by the crystallographic texture and the resulting stress state relative to the c-axis of the hexagonal lattice [50,51]. In the compression zone (CZ), the applied compressive stress induces an extension along the c-axis, which strongly favors the activation of $\{10\bar{1}2\}$ extension twinning due to its low critical resolved shear stress (CRSS) [54]. As a result, twinning becomes the dominant deformation mechanism in this region and accommodates a significant portion of the strain imposed. This behavior is further enhanced by high Schmid factor values for twinning in grains with basal texture, leading to widespread twin formation across the CZ [55].

In contrast, in the tension zone (TZ), tension along the wire axis contracts rather than extends the c-axis, so $\{10\bar{1}2\}$ extension twinning is not likely to be activated irrespective of the resolved shear stress magnitude. Consequently, deformation is predominantly accommodated by dislocation slip, primarily basal slip. Although a limited amount of twinning may occur in specific grains due to secondary stress components (e.g., transverse tensile stresses) according to Shi and Zhang [56], it does not represent the primary deformation mechanism.

In our previous study [11] for pure Mg wires prepared via a similar extrusion process as described in Section 2.1 we observed similar bending behavior as in the present study. However, unlike in the previous study, where residual twins were observed even in the TZ in the CD map after bending and bending in the opposite direction of a pure Mg wire, the present Mg-Zn alloys depict a completely reversible twinning-detwinning mechanism after one cycle of bending. Dilute Zn alloying in these thin wires (0.6 wt% Zn studied in detail, see Fig. 4e–g) therefore does not appear to suppress detwinning. The higher CRSS for twin growth in solid-solution Mg-Zn [54] may play a role. However, this alone cannot explain the observed behavior, since many grains in Fig. 4e are fully consumed by a single twin and still revert after one cycle in Fig. 4f (see also Supplement for respective IPF maps). We plan to study this in detail in our future work.

For pure Mg, KAM maps in both cases of the tensile test deformation show higher values due to the applied uniaxial tensile stress, with twinning observed in the outer parts of the wire. The Mg-0.6Zn wire showed no twinning in the post-tensile test, being terminated in the plastic deformation region after 2 and 20 % deformation. Both of these differences could be partially due to the more pronounced $\{10\bar{1}0\}$ PF texture for Zn-containing magnesium wires, compared to the pure Mg wire. Moreover, no twinning after the tensile test is also probably affected by Zn-grain boundary segregation observed in our previous study [18], and by solid solution strengthening, where Li et al. [54] observed a higher CRSS for twin growth and basal slip for a solid solution-strengthened Mg-Zn alloy compared to pure Mg.

The bending of the wires in this study was conducted solely to observe the qualitative behavior of Mg-Zn wires and to compare the two compositions, which turned out not to alter the bending behavior. In the near future, we plan to do more advanced experiments on the bending of the wires, such as dynamic mechanical analysis (DMA) on wires during immersion in, for instance, SBF, to accurately assess their potential for use as strands in reconstructive hand surgery. Even though the bending behavior after a few bending cycles is the same for both compositions, effects such as grain boundary segregation and texture could influence the mechanism and longevity of the wire during cyclic loading or bending, even more so in the case of a corrosion medium being implemented during the testing.

### 4.3. Degradation behavior

The corrosion behavior of extruded Mg-Zn wires in SBF was evaluated based on the corrosion parameters obtained from the cathodic polarization curves. The cathodic polarization curves can be divided into two distinct regions. Close to the OCP, the cathodic response is relatively linear on a semi-logarithmic scale and can be approximated by Tafel behavior. This region is associated with the reduction of water accompanied by hydrogen evolution and was therefore used for the determination of corrosion parameters. At more negative potentials, approximately below −2.1 V, a noticeable change in slope is observed. This transition indicates a change in the kinetic regime of the hydrogen evolution reaction. In this region, the cathodic process becomes increasingly influenced by factors such as gas evolution and surface film formation, resulting in deviation from activation-controlled behavior.

The anodic branches exhibited a characteristic breakdown around −1.5 V, followed by a sharp increase in anodic current density, indicating the onset of localized corrosion. This behavior is attributed to the breakdown of the primary surface film, predominantly composed of $Mg(OH)_2$. In combination with the negative difference effect, the anodic branch deviates from the ideal Tafel behavior, making it unsuitable for reliable extrapolation and determination of corrosion parameters [57,58]. Hence, corrosion parameters were assessed from the cathodic branches. However, the obtained corrosion parameters should be interpreted with caution. Mg corrosion in SBF does not fulfill the assumption of ideal Tafel kinetics. This is further supported by relatively high Tafel slopes (240–330 mV/dec), indicating mixed control of the cathodic reaction involving hydrogen evolution and interactions with the surface film [59].

The corrosion parameters show there was no significant difference in corrosion behavior among the investigated extruded Mg-Zn wires. This indicates that the corrosion behavior of extruded wires is dominated by Mg dissolution rather than alloying effect. This conclusion is consistent with the microstructural analysis, which revealed no substantial differences among the extruded wires. This contrasts with cast Mg-Zn alloys, where increasing Zn content typically improves corrosion resistance while Zn remains in solid solution; once the solubility limit is exceeded, $Mg_xZn_y$ precipitates act as cathodic sites and accelerate matrix dissolution, reversing the trend [6,60].

Compared to the literature data [60–62], the investigated alloys exhibited more negative corrosion potentials and higher corrosion current densities. This discrepancy can be primarily attributed to the surface treatment prior to the PDP measurement. Chemical etching in nital solution effectively removes the native $MgO/Mg(OH)_2$ layer and exposes the active surface. This film partially protects magnesium and can control corrosion processes. Its breakdown or removal leads to direct exposure of metallic surface and accelerated corrosion kinetics [63,64]. This interpretation is supported by comparative measurements performed on the mechanically ground samples, which exhibited significantly lower corrosion current densities comparable to values reported in literature.

Additional contributions may arise from the underestimation of the real surface area of the wire. The surface of the wires was approximated as a smooth cylindrical geometry. Nevertheless, the μCT analysis showed, that the actual effective area can be larger. Consequently, the calculated current densities may be systematically overestimated.

Therefore, the corrosion rates obtained from the PDP measurements should be considered primarily suitable for relative comparison between the alloys rather than for determining the absolute corrosion rates.

The results discussed in Section 3.2 on the analysis of the corrosion products in Fig. 7 show that a correlative analysis of the corrosion products after biodegradation of Mg-Zn wires, combined with bulk samples XRD analysis, is necessary to sufficiently identify present corrosion products. However, it should be noted that the bulk and wire samples differ in grain size, which may influence corrosion-product formation. The bulk samples were used as a proxy for the wires, and comparative surface analyses (SEM, EDS) confirmed the formation of similar corrosion products on both sample types.

Moreover, observed apatites (hydroxyapatite and B-type carbonated apatite) display autofluorescence under UV light using the ET GFP filter, likely due to carbonate impurities, as discussed by Gonzalez et al. [65]. In our previous study, we also observed different corrosion products after immersion of the Mg-Zn-Mn alloy in modified DMEM, which displayed varying Ca:P ratios and intensities using the ET GFP filter [26]. We suggested that these intensities depend on the presence of different types of apatites. The findings in this study further support our claims.

The mechanical integrity of Mg-0.6Zn wires deteriorated rapidly upon immersion in SBF (Fig. 9a–c), exhibiting severe localized pitting that substantially reduced the load-bearing cross-section within 3 days (Fig. 9f) and likely culminated in complete dissolution by day 7 in agreement with the progressive matrix consumption demonstrated by XRD analysis (Section 3.2). These observations are consistent with the degradation mechanisms described by Song et al. [66], who reported that the mechanical properties of extruded Mg-2Zn-0.8Zr decreased rapidly during the initial immersion period in SBF as a consequence of stress concentration at corrosion pits. Whereas their bulk specimens retained measurable mechanical integrity throughout 28 days of immersion, this discrepancy is most likely attributable to the larger diameter of their specimens (4 mm) relative to the wires examined here (0.28 mm), in which even shallow pits represent a significant fraction of the cross-section.

The in vitro degradation of Mg alloys is governed by numerous factors, including the immersion medium, buffering system, fluid flow, and loading conditions; nevertheless, the *in vivo* degradation mechanisms remain incompletely understood and cannot be fully replicated in vitro [67]. Although the composition of SBF approximates that of blood plasma and is widely employed for in vitro degradation measurements [66–69], it remains a coarse approximation of the *in vivo* environment [70]. In the present study, the rapid degradation and the rise in pH above physiological values (to 9.5) indicate failure of the HCl/Tris buffering system. Furthermore, the HCl/Tris buffer may itself accelerate corrosion by increasing $Cl^-$ concentration, consuming $OH^-$ ions, or ions complexation, and thereby disrupting the formation of the protective surface layer [68]. SBF with Tris buffer nonetheless retains practical advantages as a screening medium, namely its compositional simplicity and the accelerated degradation kinetics that enable rapid comparative ranking of alloy compositions; for fine wires, however, the severity and rate of corrosion proved excessive, as even modest pitting compromises the entire load-bearing cross-section.

To overcome these limitations and to assess degradation under more physiologically relevant conditions, a second series of immersion tests on Mg-0.6Zn wires was conducted in DMEM supplemented with FBS under a 5 % $CO_2$ atmosphere at 37 °C, with all other immersion and tensile testing conditions held constant. Although experimentally more demanding, this system provides physiologically relevant buffering and is considered a closer approximation of the *in vivo* environment. Serum-supplemented cell culture media under $CO_2/HCO_3^-$ buffering reproduce both the degradation rate and the composition of the degradation layer observed *in vivo* more closely than simple salt solutions, and are therefore recommended for studies of degradation behavior and mechanisms [67], although they remain an approximation rather than a direct replica of the *in vivo* environment. A markedly different corrosion behavior was observed in DMEM + FBS. After 3 days of immersion, the corrosion layer was more compact and exhibited considerably less pitting than that formed in SBF over the same period (Fig. 9e), which is reflected in the higher retained mechanical properties shown in Fig. 9a–c and Table 5. After 7 days of immersion, $F_M$ and $F_{02}$ decreased by only 13 % and 18 %, respectively, relative to the reference state (Table 5, 0 days), whereas $A_t$ – the most sensitive indicator of degradation – declined by 68 %. The $CO_2/HCO_3^-$ buffering system also performed more effectively, with the pH increasing from an initial value of 7.4 to 7.9 after 7 days of immersion. These findings are consistent with the lower corrosion rates reported for DMEM relative to SBF, attributed to the formation of a more compact precipitate layer, the adsorption of amino acids, and the absence of Tris-HCl [68], and with the reported reduction of Mg degradation upon supplementation of cell culture media with FBS [67]. The immersion series was limited to 7 days and therefore describes only the early stage of degradation. Longer immersion periods, corresponding to the several weeks relevant for bone healing, were not covered by the present experiments.

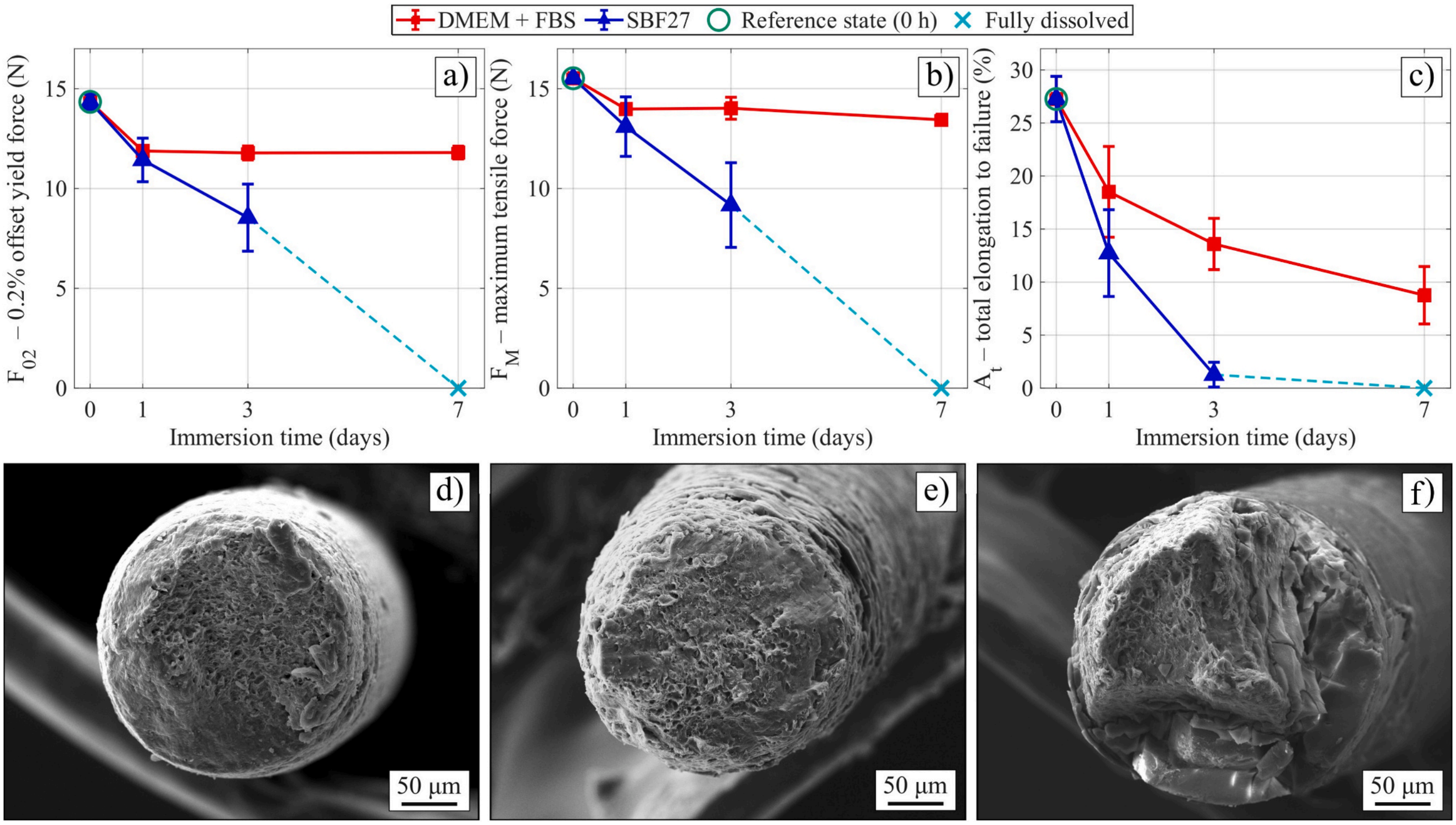

**Fig. 9.** Effect of immersion in SBF and DMEM + FBS at 37 °C on the tensile response of nital-pickled Mg-0.6Zn wires after 1, 3, and 7 days, with corresponding fracture-surface morphology. (a) 0.2 % offset yield force $F_{02}$; (b) maximum tensile force $F_M$; (c) total elongation to failure $A_t$. The cyan cross at day 7 in the SBF series denotes complete dissolution of the specimens, precluding mechanical testing. Values are given in Table 5. (d)–(f) SEM images (SE signal) of representative tensile fracture surfaces: d) reference state prior to immersion; (e) after 3 days of immersion in DMEM + FBS at 37 °C; (f) after 3 days of immersion in SBF for 37 °C. (For interpretation of the references to colour in this figure legend, the reader is referred to the web version of this article.)

### 4.4. Clinical potential and optimization of dilute Mg-Zn wires

For pediatric applications, where the long-term effects of foreign materials on a developing organism are a major concern, biocompatibility is critical. For this reason, the only intentional alloying addition in this system is zinc, while rare-earth additions were deliberately avoided owing to the limited research on their long-term biological impact [10,71]. In their extruded state, these dilute Mg-Zn wires provide a conservative mechanical and biological baseline, demonstrating adequate strength and ductility for wire processing and knotting. Building upon this solid foundation, the mechanical and degradation performance can be further optimized by microalloying with well-tolerated elements such as Mn, Sr, Ca [27,72].

The degradation rate remains the principal limitation, and it is aggravated by the small wire diameter, since even shallow pits remove a substantial fraction of the load-bearing cross-section and thus cause a disproportionate loss of ductility, a behavior also reported for other thin Mg-alloy wires [15,18]. In addition to metallurgical modifications, this sensitivity to localized corrosion can be addressed through surface engineering. Protective treatments such as micro-arc oxidation [73], biodegradable polymer [18], or composite duplex coatings have demonstrated effective results in lowering initial degradation rate [74]. Overall, while the present wires fulfill the basic mechanical requirements for low-load fixation, their primary value lies in serving as a stable baseline that can be systematically enhanced using microalloying, processing, and coating methodologies.

## 5. Conclusions

Thin biodegradable Mg–Zn wires with Zn contents of 0.4, 0.6, 0.8, and 1.5 wt% were prepared by single-step hot direct extrusion and evaluated with respect to microstructure, mechanical performance, bending deformation, and in vitro degradation. The main outcomes of this study can be summarized as follows:

1. Zn content in the range of 0.4–1.5 wt% did not substantially affect grain size, ultimate tensile strength, ductility, or bending behavior (tested for Mg-0.6Zn and Mg-1.5Zn) of the 290–300 μm-thick wires. Its effect was limited to the yielding behavior, where the 0.2 % proof stress was higher for Mg-0.4Zn and Mg-0.6Zn (229–233 MPa) than for Mg-0.8Zn and Mg-1.5Zn (200–207 MPa). This was related to the sharp yield point, followed by a post-yield stress drop, most pronounced for Mg-0.4Zn and Mg-0.6Zn wires.
2. Qualitative analysis revealed that the bending deformation was governed mainly by extrusion texture. The neutral zone shifted toward the tensile side, while tensile twinning occurred predominantly in the compression zone and was largely reversed during straightening by detwinning, preserving twinning–detwinning-assisted bending plasticity in the measured Mg-0.6Zn and Mg-1.5Zn wires.
3. Promising mechanical properties were achieved despite the use of alloys processed through a standard recycling route and despite the presence of MgO-based impurities and surface imperfections. The reported properties therefore represent a conservative lower estimate of the performance attainable with higher-purity feedstock and improved quality control.
4. SBF provided rapid corrosion screening but caused severe localized degradation and fast loss of mechanical integrity of thin Mg-0.6Zn wires. In DMEM + FBS, the wires retained 87 % of their maximum tensile force but only 32 % of their elongation after 7 days. More complex media, such as DMEM + FBS under a $CO_2$ atmosphere, are therefore needed to obtain application-relevant information for biodegradable implant performance, a motivation for follow-up studies.

## Declaration of generative AI and AI-assisted technologies in the manuscript preparation process

During the preparation and revision of this work, the authors used GPT-5.6 Sol by OpenAI and Claude Opus 5.5 by Anthropic to assist with language polishing, grammar correction, and improvements to textual clarity and flow. After using these tools, the authors reviewed and edited the content as needed and take full responsibility for the content of the published article.

## CRediT authorship contribution statement

**Jiří Ryjáček:** Writing – review & editing, Writing – original draft, Visualization, Validation, Project administration, Methodology, Investigation, Formal analysis, Conceptualization. **Leonard Hlodák:** Writing – review & editing, Writing – original draft, Visualization, Validation, Methodology, Investigation, Formal analysis, Data curation. **Jiří Liška:** Writing – review & editing, Writing – original draft, Validation, Methodology, Investigation. **Jan Pinc:** Writing – review & editing, Writing – original draft, Investigation. **Tomáš Herma:** Resources, Methodology, Conceptualization. **Karel Tesař:** Conceptualization, Funding acquisition, Methodology, Project administration, Resources, Supervision, Validation, Writing – review & editing.

## Declaration of competing interest

The authors declare that they have no known competing financial interests or personal relationships that could have appeared to influence the work reported in this paper.

## Acknowledgments

J.R., L.H., J.L., and K.T. were supported by the Czech Science Foundation (GAČR), Junior Star project No. 25-17788M. L.H., J.P., and K.T. acknowledge the support provided by the Ferroic Multifunctionalities project, supported by the Ministry of Education, Youth, and Sports of the Czech Republic; Project No. CZ.02.01.01/00/22_008/0004591, co-funded by the European Union. J.L. was also supported by the Student Grant Competition of the Czech Technical University in Prague, grant No. SGS25/171/OHK4/3T/14. All authors further acknowledge project No. CZ.10.02.01/00/25_082/0001369, Resorbable metal implants for reconstructive hand surgery, based at the Děčín workplace of the Czech Technical University in Prague, supported by the Ministry of the Environment of the Czech Republic and co-funded by the European Union through the Operational Programme Just Transition. The authors sincerely thank Karel Trojan, Martin Libich, and Jan Drahokoupil for their technical assistance and valuable discussions.

## Appendix A. Supplementary data

Supplementary data to this article can be found online at https://doi.org/10.1016/j.matdes.2026.117246.

## Data availability

Data are available online on Zenodo (https://doi.org/10.5281/zenodo.19917969) and on request.